# The Relative Angle Distribution Function in the Langevin Theory of Dilute Dipoles

Robert D. Nielsen

ExxonMobil Research and Engineering Co., Clinton Township, 1545 Route 22 East, Annandale, NJ 08801

robert.nielsen@exxonmobil.com

**Abstract**

The Langevin theory of the polarization of a dilute collection of dipoles by an external field is often included in introductory solid state physics and physical chemistry curricula. The average polarization is calculated assuming the dipoles are in thermal equilibrium with a heat bath. The heart of the polarization calculation is a derivation of the average dipole-field projection, whose dependence on the external field is given by the Langevin function. The Langevin problem is revisited, here, and the average projection of any given dipole onto any other dipole from the collection is derived in terms of the Langevin function. A simple expression is obtained for the underlying dipole-dipole angular distribution function.

**I. Introduction**

A single magnetic dipole $\vec{\mu}$ in an external magnetic field $\vec{H}$ has a potential energy: $V = -\vec{\mu} \cdot \vec{H} = -|\vec{\mu}| \cdot |\vec{H}| \cos(\theta).$[1] While formulating a theory of magnetism, Langevin considered a collection of dipoles in an external magnetic field.[2] The concentration of the dipoles was assumed to be sufficiently diluted that dipole-dipole interactions could be neglected, leaving only the sum over the individual dipole-field potential energies for the total energy. Langevin developed the equilibrium average value of the dipole projection on to the external field, $\langle \cos(\theta) \rangle$, by assuming that the dipoles were in contact with a heat bath. The distribution function, which allows the equilibrium averages to be calculated, is the Boltzmann distribution:

$$\frac{e^{-V/kT}}{Z} = \frac{e^{F \cdot \cos\theta}}{Z_L(F)}$$

where $F = |\vec{\mu}| \cdot |\vec{H}|/kT$, and

$$Z_L(F) = \int_{\phi=0}^{2\pi} \int_{\theta=0}^{\pi} e^{F \cdot \cos\theta} \sin(\theta) \cdot d\theta \cdot d\phi = \int_{\phi=0}^{2\pi} \int_{\cos\theta=-1}^{1} e^{F \cdot \cos\theta} d\cos\theta \cdot d\phi = 4 \cdot \pi \cdot \frac{\sinh(F)}{F}.$$

$Z_L(F)$ is the partition function.[3,4] $k$ and $T$ are Boltzmann's constant and the temperature of the heat bath.

The equilibrium averages are:

$$\langle \cos^n(\theta) \rangle = \frac{1}{Z_L} \cdot \frac{d^n Z_L(F)}{dF^n} = L_n(F)$$

Where $L_n(F)$ are the $n^{th}$ order Langevin functions. $L_1 = \langle \cos(\theta) \rangle$ is often simply denoted by $L$.[1,3]

We now ask, for the same system of dipoles, what is average projection of any given dipole onto any other dipole from the collection. We also ask, what is the distribution function for this relative projection in terms of the value of the external field? Section II develops the relative angle distribution function and averages. Section III shows the relative angle distribution function that is generated numerically from Monte Carlo calculations with some trial values of $F$, for comparison.

**II. Relative angle distribution and averages**

Figure 1 shows the relative orientation of two dipoles, labeled 1 and 2. The relative angle between the two dipoles is denoted by $\gamma$. The angles that define each dipole's projection onto the external Z-axis are given by $\theta'$ and $\theta''$ for dipoles 1 and 2 respectively. The Z-axis will be taken as the direction of the external field $\vec{H}$, so that $\theta'$ and $\theta''$ are the angles that enter the expression for the potential energy of dipoles 1 and 2.

The angle $\Delta_\gamma$ is given the subscript $\gamma$ because it is opposite the angle $\gamma$ on the spherical triangle formed from the two dipoles and the Z axis, see figure 1. $\Delta_{\theta'}$ and $\Delta_{\theta''}$ are defined likewise.

Because the dipoles are dilute, the average $\langle \cos^n(\gamma) \rangle$ can be expressed in terms of the dipole-field Boltzmann distributions of the individual dipoles 1 and 2.

$$\langle \cos^n(\gamma) \rangle = \int_{\phi'=0}^{2\pi} \int_{\cos\theta'=-1}^{1} \int_{\phi''=0}^{2\pi} \int_{\cos\theta''=-1}^{1} \cos^n(\gamma) \cdot \frac{e^{F \cdot \cos\theta''}}{Z_L(F)} \cdot d\cos\theta'' \cdot d\phi'' \cdot \frac{e^{F \cdot \cos\theta'}}{Z_L(F)} \cdot d\cos\theta' \cdot d\phi' \quad (1)$$

The two integrals over the two angles $\phi'$ and $\phi''$ can be replaced with a single integral over the relative angle $\Delta_\gamma = \phi' - \phi''$ because $\cos(\gamma)$ is periodic in both variables $\phi'$ and $\phi''$, and the integration ranges extend from 0 to $2\pi$.

Equation (1) then becomes:

$$\langle \cos^n(\gamma) \rangle = 2\pi \cdot \int_{\Delta_\gamma=0}^{2\pi} \int_{\cos\theta'=-1}^{1} \int_{\cos\theta''=-1}^{1} \cos^n(\gamma) \cdot \frac{e^{F \cdot \cos\theta'}}{Z_L(F)} \cdot \frac{e^{F \cdot \cos\theta''}}{Z_L(F)} \cdot d\cos\theta'' \cdot d\cos\theta' \cdot d\Delta_\gamma \quad (2)$$

The angle addition formula,

$$\cos(\gamma) = \cos(\theta') \cdot \cos(\theta'') + \sin(\theta') \cdot \sin(\theta'') \cdot \cos(\Delta_\gamma), \quad (3)$$

from spherical trigonometry gives the dependence of $\cos(\gamma)$ on the integration variables.[5]

Direct calculation of the integral (2) with $n=1$ and $n=2$, for example, gives:

$$\langle \cos(\gamma) \rangle = \langle \cos(\theta') \rangle \cdot \langle \cos(\theta'') \rangle = (L_1(F))^2$$

and

$$\langle \cos^2(\gamma) \rangle = \langle \cos^2(\theta') \rangle \cdot \langle \cos^2(\theta'') \rangle + \frac{1}{2} \cdot (1 - \langle \cos^2(\theta') \rangle) \cdot (1 - \langle \cos^2(\theta'') \rangle)$$

$$= \frac{3}{2} \cdot (L_2(F))^2 - L_2(F) + \frac{1}{2}$$

While any of the average values $\langle \cos^n(\gamma) \rangle$ can be calculated in this manner, by expanding $\cos^n(\gamma)$ in the integrand using the angle addition formula, the underlying distribution of $\cos(\gamma)$ that governs the averages is not transparent. A distribution function, $\rho(\cos(\gamma))$, is sought, such that:

$$\langle \cos^n(\gamma) \rangle = \int_{\cos\gamma=-1}^{1} \cos^n(\gamma) \cdot \rho(\cos(\gamma)) \cdot \mathrm{d}\cos(\gamma).$$

To establish the distribution $\rho(\cos(\gamma))$, a change of variables is made in equation (2) . The variables in equation (2) consist of two sides (arcs) of a spherical triangle and the intervening vertex angle (see figure 1). The integral (2) can be re-expressed, in general, in terms of any two sides of the spherical triangle and their vertex angle. So, for example, the following transformation is possible:

$$\{\cos(\theta'), \cos(\theta''), \Delta_\gamma\} \to \{\cos(\theta'), \cos(\gamma), \Delta_{\theta''}\} \qquad (4)$$

That the Jacobian is unity for this transformation maybe be verified analytically by calculating $\partial\cos(\gamma)/\partial\Delta_\gamma$, $\partial\cos(\gamma)/\partial\cos(\theta'')$, $\partial\Delta_{\theta''}/\partial\Delta_\gamma$ and $\partial\Delta_{\theta''}/\partial\cos(\theta'')$ with the aid of the addition formula: $\cos(\theta'') = \cos(\theta') \cdot \cos(\gamma) + \sin(\theta') \cdot \sin(\gamma) \cdot \cos(\Delta_{\theta''})$, the law of sines for spherical triangles: $\sin(\theta'')/\sin(\Delta_{\theta''}) = \sin(\gamma)/\sin(\Delta_\gamma)$, and the auxillary formula: $\sin(\gamma) \cdot \cos(\Delta_{\theta_2}) = \partial\cos(\gamma)/\partial\theta'$ ,from spherical trigonometry.[5]

With the transformation (4), equation (2) becomes:

$$\left\langle \cos^n(\gamma) \right\rangle = \frac{2 \cdot \pi}{\left(Z(F)\right)^2} \cdot \int\limits_{\cos\gamma=-1}^{1} \int\limits_{\Delta_{\theta'}=0}^{2\pi} \int\limits_{\cos\theta'=-1}^{1} \cos^n(\gamma) \cdot e^{F\left(\cos(\theta')\cdot(1+\cos(\gamma))+\sin(\theta')\cdot\sin(\gamma)\cdot\cos(\Delta_{\theta'})\right)} \cdot d\cos\theta' \cdot d\Delta_{\theta''} \cdot d\cos\gamma$$

Dropping the subscript and superscript on the angles, $\rho(\cos(\gamma))$ is identified as:

$$\rho(\cos(\gamma)) = \frac{2 \cdot \pi}{\left(Z_L(F)\right)^2} \cdot \int\limits_{\cos\theta=-1}^{1} \int\limits_{\Delta=0}^{2\pi} e^{F\left(\cos(\theta)\cdot(1+\cos(\gamma))+\sin(\theta)\cdot\sin(\gamma)\cdot\cos(\Delta)\right)} \cdot d\Delta \cdot d\cos\theta \qquad (5)$$

The integrand in equation (5) is simplified by the following change of parameters:

$$a = F \cdot (\cos(\gamma)+1) \quad , \quad b = F \cdot \sin(\gamma) \quad , \quad c = \sqrt{a^2+b^2} = F\sqrt{2 \cdot (1+\cos(\gamma))} \quad ,$$

$$\sin(\alpha) = b/c = F \cdot \sin(\gamma)/c \quad , \quad \cos(\alpha) = a/c = \sqrt{c^2-b^2}/c = F \cdot (\cos(\gamma)+1)/c$$

The equation (5) is transformed to:

$$\rho(\cos(\gamma)) = \frac{2 \cdot \pi}{\left(Z_L(F)\right)^2} \cdot \int\limits_{\cos\theta=-1}^{1} \int\limits_{\Delta=0}^{2\pi} e^{c \cdot (\cos(\alpha)\cdot\cos(\theta)+\sin(\alpha)\cdot\sin(\theta)\cdot\cos(\Delta))} \cdot d\Delta \cdot d\cos\theta \qquad (6)$$

The cosine addition formula, (3), and an angular transformation, analogous to (4), allows the argument of the exponential in the integrand to be written as a single cosine. The expression for the distribution function (6) is integrated to give:

$$\rho(\cos(\gamma)) = \frac{1}{2} \cdot \left(\frac{4 \cdot \pi}{Z_L(F)}\right)^2 \cdot \frac{\sinh\left(F\sqrt{2 \cdot (1+\cos(\gamma))}\right)}{F\sqrt{2 \cdot (1+\cos(\gamma))}}$$

$$= 2 \cdot \pi \cdot \frac{Z_L\left(F\sqrt{2 \cdot (1+\cos(\gamma))}\right)}{\left(Z_L(F)\right)^2} \qquad (7)$$

The normalization of the distribution (7) is verified by the change of variables:

$$u = F\sqrt{2 \cdot (1+\cos(\gamma))}.$$

$$\int_{\cos\gamma=-1}^{1} \rho(\cos(\gamma)) \cdot d\cos(\gamma) = \frac{1}{2} \cdot \left(\frac{4 \cdot \pi}{Z_L(F)}\right)^2 \cdot \int_{\cos\gamma=-1}^{1} \frac{\sinh\left(F\sqrt{2 \cdot (1+\cos(\gamma))}\right)}{F\sqrt{2 \cdot (1+\cos(\gamma))}} \cdot d\cos(\gamma)$$

$$= \frac{1}{2} \cdot \left(\frac{4 \cdot \pi}{Z_L(F)}\right)^2 \cdot \frac{1}{F^2} \int_0^{2F} \sinh(u) \cdot du = \frac{1}{2} \cdot \left(\frac{4 \cdot \pi}{Z_L(F)}\right)^2 \cdot \frac{\cosh(2 \cdot F) - 1}{F^2}$$

$$= \frac{1}{2} \cdot \left(\frac{4 \cdot \pi}{Z_L(F)}\right)^2 \cdot 2 \cdot \left(\frac{\sinh(F)}{F}\right)^2 = 1$$

The averages, $\langle \cos^n(\gamma) \rangle$, can be expressed, likewise by a change of variables, as:

$$\langle \cos^n(\gamma) \rangle = \frac{1}{2 \cdot \sinh^2(F)} \int_0^{2F} \left(1 - \frac{u^2}{2 \cdot F^2}\right)^n \cdot \sinh(u) \cdot du$$

### III. Monte Carlo

Monte Carlo provides a way to numerically test the distribution function (7). Monte Carlo numerically generates configurations of dipoles in an external field that are consistent with thermal equilibrium ( Bolztmann statistics ). The input to the Monte Carlo calculation here is a set of 1000 dipoles with randomly assigned orientations. Each Monte Carlo cycle refines the orientations of all 1000 dipoles by making random changes to the individual dipole orientations, one at a time. If a given dipole's energy $(V = -F\cos(\theta))$ is decreased or remains the same as a result of the random re-orientation, the new orientation is kept and replaces the original orientation. If the random re-orientation of a dipole leads to an increase in energy, the new orientation is not always accepted. A move that increases the energy is kept with a frequency that is dictated by the Boltzmann weighting of the energy difference: $e^{-(V_{new} - V_{old})/kT}$. In other words, larger changes in energy are accepted less frequently than smaller energy changes in manner that is consistent with thermal equilibrium. An overview of Monte Carlo is available in standard texts, where sample codes are given.[6, 7] A Monte Carlo simulation was run on a set of 1000 dipoles for each of the values: $F = 1/5$, 1, 2, and 5. The last 400 Monte Carlo cycles out of a 5000 cycle trajectory were used to compile statistics of the dipole orientations for each value of $F$. Figure 2 shows normalized histograms of the values of $\cos(\theta)$ (figure 2, left panels) and $\cos(\gamma)$ (figure 2, right panels) from the Monte Carlo dipole configurations. The values of $F$ increase from top to bottom. The solid lines are the Boltzmann distribution (figure 2, left panels), and the relative angle distribution function (7) (figure 2, right panels). The Monte Carlo results numerically confirm the analytic relative angle distribution (7) derived in section II.

**IV. Discussion**

A demonstration of the relative angle distribution function for dipoles in the dilute limit serves two purposes. Firstly, the derivation can be used as a follow up exercise to the standard Langevin problem. The calculation of the relative angle distribution is slightly more challenging than the calculation of the average polarization. Furthermore, the problem introduces the idea of relative vs. external orientational order, which foreshadows the introduction of an angular distribution function in condensed phase statistics. The following qualitative question might be posed, for example, to help explore the difference between the relative and external order: Why, in figure 2, does the relative angle distribution function appear to "lag behind" the Boltzmann distribution function in its dependence on the external field?

Secondly, the analytical expression for the distribution function derived in the dilute limit (equation (7)) is useful for comparison with the angular distribution function that arises when dipole-dipole interactions are present at higher densities. The general angular distribution function at high density reflects both the influence of the externally applied field, as well as dipole-dipole interactions. The collective dipole-dipole interactions are not easily described by simple analytic formulae because dipoles form phases and exhibit long range order that involves the participation of many dipoles.[7] A common method of grasping the structure of the dipolar phases visually is to choose a representative dipole and then record the angular distribution of all other dipoles that are at some fixed distance from the central dipole. This procedure is repeated for multiple representative dipoles and distances, and statistics of the relative angle distribution are compiled. One way of testing whether the field-dipole interaction dominates the dipole order is to compare the relative angle distribution function that is observed to the dilute limit given by equation (7).

**Acknowledgments**

The author wishes to thank REU student Field N. Cady, and Dr. Bruce H. Robinson for careful reading of the manuscript before submission.

**Figure Captions:**

**Figure 1:** The geometry of two dipoles (dark arrows). $\gamma$ is the relative angle between dipoles 1 and 2. $\theta'$ and $\theta''$ are the projection angles of dipoles 1 and 2 with respect to the Z-axis. $\Delta_\gamma$, $\Delta_{\theta'}$, and $\Delta_{\theta''}$ are the vertex angles of the spherical triangle formed by the two dipoles and the Z-axis.

**Figure 2:** Histograms compiled from Monte Carlo simulations with 1000 dipoles in an external field with $F = 1/5, 1, 2,$ and $5$ (increasing from top to bottom).
Left panels: Normalized histogram of $\cos(\theta)$ from Monte Carlo (bars) overlaid with the Boltzmann distribution (solid lines).
Right panels: Normalized histogram of $\cos(\gamma)$ from Monte Carlo (bars) overlaid with the relative angle distribution (equation (7),text)

**Figure 1**

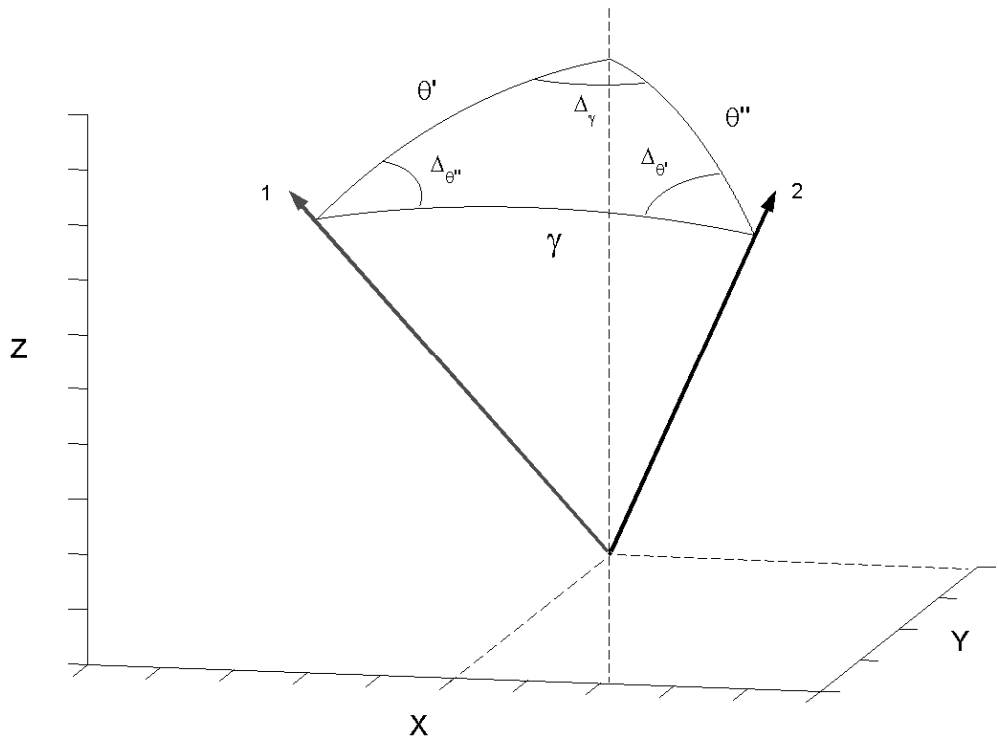

**Figure 2**

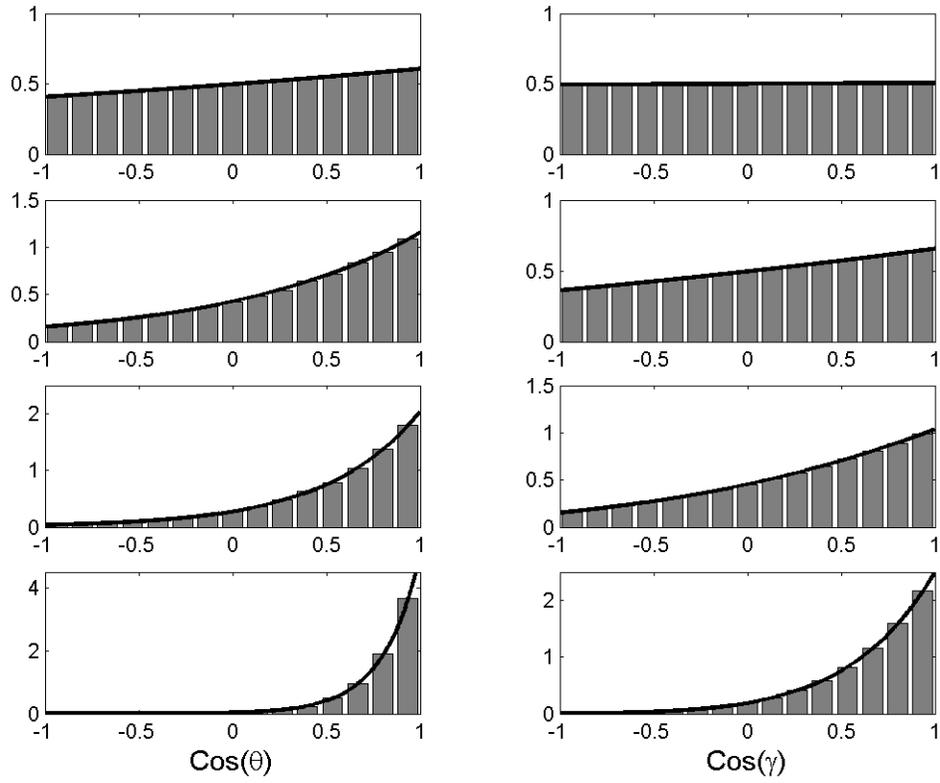